\documentclass[
    ,final            
  ]
  {aipproc}

\layoutstyle{8x11double}

\begin{document}

\title{Specific adhesion of peptides on semiconductor surfaces in
experiment and simulation}



\author{Karsten Goede}{
  address={Institut f{\"u}r Experimentelle Physik II, Universit{\"a}t Leipzig, 
Linn\'{e}stra{\ss}e 5, 04103 Leipzig, Germany}
}

\author{Michael Bachmann}{
  address={Institut f{\"u}r Theoretische Physik, Universit{\"a}t Leipzig, 
Augustusplatz 10-11, 04109 Leipzig, Germany}
}

\author{Wolfhard Janke}{
  address={Institut f{\"u}r Theoretische Physik, Universit{\"a}t Leipzig, 
Augustusplatz 10-11, 04109 Leipzig, Germany}
}

\author{Marius Grundmann}{
  address={Institut f{\"u}r Experimentelle Physik II, Universit{\"a}t Leipzig, 
Linn\'{e}stra{\ss}e 5, 04103 Leipzig, Germany}
}

\begin{abstract}
We report on self-assembly, clustering, and conformational phases
of peptides on inorganic semiconductor surfaces.
The peptide-covered surface fraction can differ by a factor of 25, depending 
mainly on surface and peptide polarity. Low 
adhesion induces large and soft clusters, which also 
have high contact angles to the surface. Direct surface adhesion of a peptide molecule competes with forming molecular aggregates which offer an overall 
reduced surface contact. 
Simulating a simple hybrid model yields a pseudophase diagram with
a rich, temperature and solvent-quality dependent variety of subphases which
are specific to the hydrophobicity and polarity of the considered substrates.
\end{abstract}

\maketitle



\section {observing peptide adhesion}

Hybrid organic-inorganic interfaces built up by specific peptide adhesion on semiconductors provide a promising model system for molecular self-assembly. Hybrid devices could prove superior to today's solutions in sensing or medicine and enable novel fields like nano-bio electronics. Yet to date, with a microscopical adhesion model still lacking, designing peptide sequences with a desired adhesion behavior is still a challenge.

Most measurements employ the small peptide AQNPSDNNTHTH. It has been bred for good adhesion on GaAs (100) \cite{belcher} and it has been shown \cite{nanolett} that its peptide adhesion coefficient (PAC) on various inorganic semiconductors ranges from 25\,\% on GaAs to 1\,\% on Si under the same conditions. 
Suitable clean and flat substrate pieces \cite{nanolett} have been exposed to a diluted watery solution of the peptide 
(1 $\mu$g/mL, pH 7.6, Tris-buffered saline).
%
\begin{figure} 
\resizebox{\columnwidth}{!}{
\includegraphics{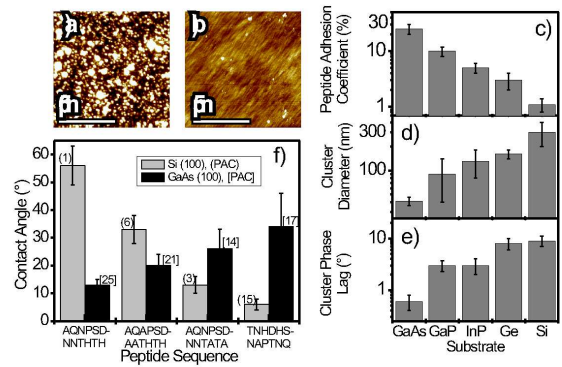}}
\caption{Peptide adhesion on semiconductors. a, b) AFM images of AQNPSDNNTHTH peptide on a) GaAs (100) (455 clusters) and b) Si (100) (16 clusters). c)-e)  AQNPSDNNTHTH adhesion and clustering on five (100) semiconductor surfaces. c) PAC, d) cluster diameter, e) cluster softness, as indicated by the phase-lag signal. f) Cluster--substrate contact angle for four different peptides on (100) surfaces of GaAs and Si, respectively. Bracket numbers denote the PAC on the respective surface.}
\label{fig:1}
\end{figure}
%
After washing with distilled water and drying in air, sample surfaces have been investigated by atomic-force microscopy (AFM) in tapping mode. PACs have been obtained from a grain analysis of each image.

The surface specificity of peptide self-assembly on semiconductor surfaces is demonstrated in Figs.~\ref{fig:1}a,b, which feature many small clusters on GaAs (100) (a) and few large clusters on Si (100) (b). (Note the different scale bars.) The less similar a surface is to that of GaAs in terms of electronegativity, the smaller is the AQNPSDNNTHTH PAC on that surface (Fig.~\ref{fig:1}c). A detailed analysis reveals that this can be ascribed to a large extent to the interplay between the dominantly polar amino-acid side chains of this peptide with the more or less polar surface \cite {nanolett}. Peptide adhesion to semiconductors is found to happen by formation of surface-specific clusters: The lower the PAC on a surface, the larger (Fig.~\ref{fig:1}d) and softer (Fig.~\ref{fig:1}e) the respective clusters become \cite{langmuir}. Since the softness of such clusters has been experimentally found to depend only weakly on their size, the substrate-specific cluster softness appears to be induced by the substrate's attractiveness to an approaching peptide molecule: Its direct adhesion on the surface competes with joining one of the already existing molecular aggregates at the surface. Aiming at the smallest possible surface energy, the outcome of this contest is system-dependent. Further evidence for the validity of this model comes from investigating the system-dependent cluster contact angle to the surface: Fig.~\ref{fig:1}f shows that this angle varies beween $5^{\circ}$ and $55^{\circ}$, depending both on peptide sequence and substrate. A large contact angle generally indicates a low attractiveness of a given substrate to a peptide molecule with a certain sequence.

When samples are prepared in peptide solutions with different pH values, the various amino-acid side chains are charged in different ways, which yields a pH dependence of the respective PAC. The adhesion impact of different conformational phases of the peptide has been studied in first measurements, with the more rigid conformations adhering worse.

\section{simulating a hybrid interface}
%
For a comprising qualitative analysis of the peptide adsorption process 
to specific substrates, 
computer simulations of simple models are extremely useful. Here, we
employ the minimalistic hybrid model with energy
$E=-n_s-s n_{\rm HH}$
on a simple-cubic lattice~\cite{bj1}, here $s$ is an effective 
solubility parameter, 
$n_s$ the number of substrate-dependent contacts with the attractive surface and 
$n_{\rm HH}$ the number of intrinsic hydrophobic nearest-neighbor contacts.
An exemplified hydrophobic-polar peptide with 103 monomers~\cite{bj1} is modeled 
as a self-avoiding chain. 
We distinguish the \emph{unspecific} substrate, where hydrophobic 
and polar monomers are equally attracted, and the specific 
\emph{hydrophobic} (like Si) and \emph{polar} (e.g., GaAs) substrates.
\begin{figure}
  \includegraphics[height=.46\textheight]{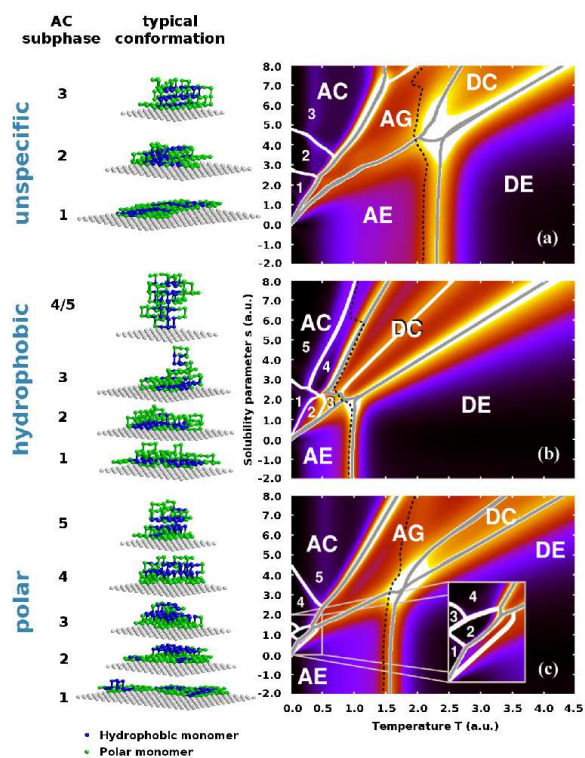}
  \caption{\label{fig:pd103}Specific heat profiles for substrates being (a) unspecifically attractive, (b) hydrophobic, and (c) polar. Also shown
are typical conformations in the AC subphases.}
\end{figure}
In our simulation, we applied a generalized variant of the multicanonical chain-growth
method~\cite{bj3}. In Fig.~\ref{fig:pd103}, the specific heat profiles as 
function of temperature $T$ and solubility $s$ are shown. 
Ridges (marked by white and gray lines) indicate conformational 
pseudophase transitions. In all cases, there is a strong first-order-like
unbinding transition between adsorbed and desorbed pseudophases. In the bulk,
the typical expanded random-coil-like conformations (DE) and the compact, 
native-like folds (DC) can be distinguished. In the adsorbed regime 
we also find expanded (AE) and compact/globular (AC, AG) phases.
Even more exciting, there is a rich substrate-dependent AC subphase structure. 
Typical conformations, also shown
in Fig.~\ref{fig:pd103}, reveal that the formation and compactness of the 
hydrophobic domains not only depend on the solvent quality (which influences, e.g., 
layering), but also on the passive, steric (hydrophobic substrate) or 
active attraction (polar substrate) of polar residues. 

Future applications require understanding of peptide adsorption mechanisms, making experimental verification as well as microscopic modeling and simulation rewarding tasks.
\begin{theacknowledgments}

We thank V.\ Gottschalch, H.\ Herrenberger, F.\ Kremer, D.\ Haines,
and A.\ Beck-Sickinger for providing substrates and etches, AFM equipment, and 
peptide synthesis, respectively.
This work is partially supported by the DFG grant
under Contract No.\ JA 483/24-1 and the JUMP supercomputer time grant No.\ hlz11
of the NIC J\"ulich.
\end{theacknowledgments}



\end{document}